# Market efficiency, informational asymmetry and pseudo-collusion of adaptively learning agents

Aleksei Pastushkov[1]

Department of Economics, Higher School of Economics, Moscow, Russia

*Abstract.* We examine the dynamics of informational efficiency in a market with asymmetrically informed, boundedly rational traders who adaptively learn optimal strategies using simple multiarmed bandit (MAB) algorithms. The strategies available to the traders have two dimensions: on the one hand, the traders must endogenously choose whether to acquire a costly information signal, on the other, they must determine how aggressively they trade by choosing the share of their wealth to be invested in the risky asset. Our study contributes to two strands of literature: the literature comparing the effects of competitive and strategic behavior on asset price efficiency under costly information as well as the actively growing literature on algorithmic tacit collusion and pseudo-collusion in financial markets. We find that for certain market environments (with low information costs) our model reproduces the results of Kyle [1989] in that the ability of traders to trade strategically leads to worse price efficiency compared to the purely competitive case. For other environments (with high information costs), on the other hand, our results show that a market with strategically acting traders can be *more* efficient than a purely competitive one. Furthermore, we obtain novel results on the ability of independently learning traders to coordinate on a pseudo-collusive behavior, leading to non-competitive pricing. Contrary to some recent contributions (see e.g. [Cartea et al. 2022]), we find that the pseudo-collusive behavior in our model is robust to a large number of agents, demonstrating that even in the setting of financial markets with a large number of independently learning traders non-competitive pricing and pseudo-collusive behavior can frequently arise.

---

[1] Contact: Aleksei Pastushkov, Higher School of Economics, Department of Economics, Pokrovsky Blvd. 11, 109028 Moscow, Russia. E-mail: apastushkov@gmail.com

**Section 1. Introduction**

The question of market efficiency under asymmetric costly information is one of the key research themes in financial economics. Starting from the impossibility theorem of [Grossman & Stiglitz, 1980], there has been a steady stream of contributions examining how information is incorporated into market prices of financial assets in various settings.

For instance, while [Grossman & Stiglitz, 1980] assume competitive behavior by traders, [Kyle 1989] argues that this assumption is implausible and studies market efficiency and endogenous information acquisition in a setting with strategic traders. He finds that strategic trading leads to diminished price efficiency compared to the competitive case and higher information costs lead to lower price efficiency. Other researchers, such as e.g. [Diamond & Verrechia, 1981] and [Verrechia 1982], relaxed the original assumption of the single all-encompassing information signal and examined how diverse, distributed among many traders pieces of information are reflected in the prices. Furthermore, [Hellwig 1982] noted an inconsistency in the model of [Grossman & Stiglitz, 1980] which has to do with the assumption that their traders *simultaneously* shape asset prices with their net demand *and* infer private information from the prices. [Hellwig 1982] proposes a dynamic model in which traders condition their demand on *previously observed* prices, thus avoiding the inconsistent assumptions. More recently, there has been research on the effects of market participants' limited attention on the adjustment of prices to new information (e.g. [Boulatov et al., 2009]) as well as on the impact of heterogeneous risk preferences and other agent characteristics ([Gong & Diao, 2022]). Apart from the latter paper, however, the theoretical literature on the informational efficiency of asset prices has tended to assume some sort of maximizing behavior on the part of the agents as well as common knowledge among them regarding each other's strategies and preferences. Arguably, however, in the context of financial markets, where the secrecy of strategies and intentions represents a source of competitive advantage, this assumption is quite strong, and a more realistic description of an individual agent's behavior may be obtained by assuming some kind of adaptive learning in an environment characterized by information asymmetries, rather than common knowledge and rational expectations.

Coincidentally, recently there has been a growing interest in examining how bounded rationality in the form of adaptive learning affects the results of some well-known economic models. A particularly active research direction has been the study of how independently learning agents may converge on collusive or pseudo-collusive behavior in various economic settings, without explicit communication. For instance, [Calvano et al., 2020] study a model of Bertrand competition between two firms using a simple reinforcement learning algorithm to set their prices. They find that the prices which the competing firms converge on deviate substantially from the competitive prices predicted by the static Bertrand model. Moreover, the supra-competitive pricing is sustained by a punishment scheme which each firm uses if its competitor attempts to deviate from the collusive price. [Klein 2021] studies tacit collusion under sequential pricing and finds that in this setting learning agents converge on collusive behavior as well. [Rocher et al., 2023] find that collusive behavior can arise even if competing firms use different learning algorithms, as long as one of the firms can endogenize the learning algorithms of the others and manipulate them in such a way that they converge to supra-competitive pricing. [Asker et al., 2024], on the other hand, find that collusive behavior does not arise if agents learn not only from experience, but also by examining counterfactuals. [Hettich 2021] models the learning of competing economic agents by deep reinforcement learning and finds that while the use of artificial neural networks speeds up the learning process, collusive behavior disappears for large oligopolies with up to 10 firms. To summarize, the results on tacitly collusive behavior seem to depend significantly on the particular market setting, the information sets available to the agents and the learning algorithms employed.

As modern financial markets are dominated by algorithmic traders, a natural question arises whether there also exist opportunities in financial markets for independently learning algorithms to converge on collusive behavior. The well-known results of [Christie & Schultz, 1994] and [Christie et al., 1994] indeed provide some empirical evidence that supra-competitive pricing in financial markets is quite common. However, as [Van den Boer et al., 2022] note, supra-competitive pricing by itself does not necessarily constitute evidence of collusion since it also requires the presence of a reward-punishment scheme, which is hard to ascertain empirically. The recent advances in machine learning algorithms and the rapid growth of computational resources, however, make it possible to study such questions with the help of multiagent simulations (see e.g. the arguments advocating this approach in [Lussange et al., 2018], [Lussange et al., 2021]). Indeed, recently there has been a number of studies employing multiagent simulation to examine whether tacit collusion arises in financial markets. For instance, [Colliard et al., 2022] model the behavior of algorithmic market makers in the presence of adverse selection by Q-learning[2] and find, counterintuitively, that higher exposure to adverse selection tends to lead the market makers to adopt more competitive pricing. [Han 2022] studies the behavior of market makers using Q-learning as well and finds that cooperative behavior without explicit communication can arise, and moreover, introducing more agents does not necessarily eliminate the presence of supra-competitive pricing. [Cont & Xiong, 2024] use deep reinforcement learning to model the learning behavior of competing market makers and find that the agents converge on supra-competitive pricing, with spread levels strictly above the competitive benchmark. [Dou et al., 2024] study a market where market makers use a fixed rule to set prices and asymmetrically informed traders use adaptive learning to determine their demands and find evidence of pseudo-collusive behavior in this setting as well.

---

[2] A type of reinforcement learning algorithm (see [Sutton & Barto, 2018] for a detailed description).

Whereas the previously mentioned papers assume that agents can condition their learning on some observed state variables (e.g. actions of their competitors in previous periods or a history of prices), [Cartea et al., 2022] argue that in the context of financial markets traders face a more difficult problem. They can observe neither the actions nor the intentions of their competitors and it is unclear whether aggregate variables, such as e.g. the equilibrium price, carry any useful information for adaptive learning. Thus, the authors model the agents' learning using algorithms of the so-called multiarmed bandit (MAB) family, which do not condition on observations of a particular "state".[3] Additionally, the authors run computational experiments utilizing several possible implementations of MAB algorithms and find that although supra-competitive pricing and pseudo-collusion[4] does arise in some settings, it seems to require agents to use the same learning algorithm, and moreover, to start learning at the same time. Furthermore, in their model supra-competitive pricing and pseudo-collusion are not robust to increasing the number of competing agents, with even a modest number of competitors (5 and above) leading to prices converging to the perfectly competitive level. These results, together with the observation of a very slow convergence of the learning algorithms to supra-competitive pricing, lead the authors to conclude that pseudo collusion is an exceedingly unlikely outcome in financial markets and therefore does not represent a regulatory concern.

A common trait of the above-mentioned models of algorithmic competition in financial markets, however, has been the assumption of an OTC market structure in which price-setting is performed by specially designated dealers. But as [Musciotto et al., 2021] note, in modern markets the line between liquidity providers and informed speculators is blurred as "[…] in most settings, market making is not institutionalized and it is freely strategically performed by […] market participants" (see [Musciotto et al., 2021]). Centralized equity and equity derivatives exchanges are a prime example of a market where this is indeed the case.

---

[3] See [Sutton & Barto, 2018] for a definition of a "state" in the context of reinforcement learning.
[4] Informally, the authors define pseudo-collusion as a situation in which traders choose individually suboptimal strategies in order to achieve a cooperative outcome, in the absence of an actual punishment scheme for deviations.

In this paper, therefore, we set a two-fold goal. First, we seek to contribute to the literature on algorithmic collusion in financial markets by constructing a model in which a moderately large number of adaptively learning traders strategically provide liquidity in a centralized market. Our aim is to investigate whether in this setting non-competitive pricing arises as well. We also check whether the non-competitive pricing is a result of what [Cartea et al., 2022] term "pseudo-collusion". To our knowledge, this is the first study that addresses these questions in the context of centralized markets in which every trader can be a strategic liquidity provider. An additional innovation pursued in this study is that our model includes a moderately large number of agents (100), whereas prior literature on algorithmic collusion in financial markets typically models interactions of less than 10, and in many cases only 2-4 agents.

Secondly, we contribute to the literature on the incorporation of private costly information into asset prices briefly reviewed at the beginning of this section. Relaxing the rational expectations and common knowledge assumptions, we seek to investigate whether the predictions of the classical model of [Kyle 1989] hold when traders adaptively learn an optimal strategy. To our knowledge, this is also the first paper to study how costly private information gets incorporated into asset prices when traders are simultaneously boundedly rational *and* strategic. The case of boundedly rational and purely competitive traders, however, has been studied by [Pastushkov 2024], whose computational model is inspired by the static model of [Grossman & Stiglitz 1980]. The study of [Pastushkov 2024] thus provides a benchmark for us: similarly to how the effects of competitive and strategic trading can be compared in the static case using the models of [Grossman & Stiglitz 1980] and [Kyle 1989], the model of [Pastushkov 2024] and our model presented here allow us to compare the effects of competitive and strategic trading on price efficiency in a *dynamic* setting with adaptively learning agents.

The rest of the paper is structured as follows. Section 2 presents the model. Section 3 reports the results of the simulations. Section 4 presents avenues for further research and concludes.

**Section 2. The model**

Our model largely follows the design proposed in [Pastushkov 2024]. The key elements of the model are presented in the following subsections.

*2.1 Assets*

As is typical for many financial agent-based models (see e.g. [LeBaron 2001], [Axtell & Farmer, 2022]), in our market there are two assets: a risk-free "bond" in infinite supply providing a fixed return of $r_f$, and a risky asset[5] providing a stochastic payoff, following the process:

$$F_{t+1} = F_t * (1 + r_r) \qquad (1)$$

where $r_r$ is a stochastic per-period change in the payoff drawn from a normal distribution ~N(μ,σ) (the values of μ and σ, together with the other model parameters are presented in Section 2.4). There are N infinitely divisible shares of the risky asset. For all simulation runs, $F_0$ is arbitrarily set to 30 monetary units (MU), to enhance the comparability with the results in [Pastushkov 2024].

*2.2 Traders*

There are M=100 traders each of whom endogenously chooses a 2-dimensional strategy. The first dimension is whether to become informed by acquiring a costly information signal at cost *C*. If the trader *i* becomes informed, their expectation of the next period's payoff of the risky asset is equal to its actual value, i.e. the information signal perfectly reveals the value of the next period's payoff:

$$E_{it}(F_{t+1}) = F_{t+1} \qquad (2)$$

If the trader chooses to become uninformed, their expectation of $F_{t+1}$ is based on the average historically observed $r_r$ up to time *t*, which by assumption is costlessly observable by all agents:

$$E_{it}(F_{t+1}) = F_t * (1 + r_{r,it}) \qquad (3)$$

where

$$r_{r,it} = \frac{\sum_0^{T=t}(r_{r,t})}{t} \qquad (4)$$

The second dimension of a strategy is how aggressively a trader invests in the risky asset given their expectation of the next period's payoff, $E_{it}(F_{t+1})$. This second dimension is what differentiates the traders in our model from those assumed in [Pastushkov 2024]. As by assumption all traders are risk-neutral in our model, a trader is willing to invest in the risky asset as long as its market price, $P^*_{t+1}$, is below the trader's subjective expectation of the next period's payoff value discounted at the risk-free rate, $E_{it}(F_{t+1})/(1+r_f)$. The traders, however, can choose what share of their wealth to commit to the purchase of the risky asset. We discretize potential choices into a set α={1, 0.75, 0.5, 0.25, 0.01}, where the discretization is necessary for technical reasons (more on this in Section 2.4). Combining the two dimensions, the set of strategies *K* available to our traders is:

$$K = \{Inf_{\alpha=1}, Inf_{\alpha=0.75}, Inf_{\alpha=0.5}, Inf_{\alpha=0.25}, Inf_{\alpha=0.01},$$

$$Uninf_{\alpha=1}, Uninf_{\alpha=0.75}, Uninf_{\alpha=0.5}, Uninf_{\alpha=0.25}, Uninf_{\alpha=0.01}\}$$

---
[5] Which can be thought of as either a stock or a derivative contract.

[Pastushkov 2024] shows that the optimality of the informed and the uninformed strategies depends on the cost of information *C*. In Section 2.3 we show that the optimality of a particular level of α for a given trader generally depends on the choices of α made by the other traders. Since by assumption our traders do not have this knowledge, this motivates modelling their behavior by adaptive learning.

Once the two strategic choices have been made by all traders, trading takes place, the equilibrium market price of the risky asset is determined, and the trader *i*'s wealth in period *t+1* is given by:

$$W_{it+1} = \alpha_{it} * (W_{it} - C) * \left(\frac{F_{t+1}}{P^*_{t+1}}\right) + (1 - \alpha_{it}) * (W_{it} - C) * (1 + r_f) \quad (5)$$

The initial levels of wealth $W_{it=0}$ for all traders are i.i.d. random draws from a uniform distribution U(0,1000]. The equilibrium market price $P^*_{t+1}$ is determined endogenously by the Walrasian auctioneer market mechanism described in the next section.

*2.3 Market mechanism*

Having determined their policies with respect to the information acquisition as well as α, the traders compute the discounted expectations of the next period's risky payoff (as described in Section 2.2) and submit their demand schedules for the risky asset to a centralized market venue. As [Boulatov & George, 2013] explain, such schedules can be thought of as continuous limits of collections of limit orders submitted by traders in real financial markets. The individual schedules are aggregated by the market venue to an aggregate demand schedule. Since the supply of the risky asset is fixed at N shares in every period, the equilibrium price $P^*_t$ is found at the intersection of the demand and supply curves. The supply of N shares can be thought of as provided by a liquidity-seeking trader who is insensitive to the "fundamental" payoff value $F_t$.

Since both the informed and the uninformed traders have a reservation price equal to their discounted expectation of the next period's risky payoff, their risky asset demand sharply goes to 0 for any price above this point. Therefore, situations are possible in which the aggregate demand is above the supply N and yet the price cannot rise any further as this is the traders' reservation price. In these cases the equilibrium price is set equal to the reservation price and each trader gets an allocation of the risky asset proportionate to their share in the total demand at this price. We demonstrate below that in the market setting described here the traders cannot choose a profit-maximizing strategy without the knowledge of the other traders' choices with respect to α.

For simplicity, assume that there are M=2 agents in the market, both informed. Further, without loss of generality, assume $r_f$=0. Then the profit of Agent A investing in the risky asset at time *t* is given by:

$$\pi_{At} = (F_t - P_t^*) * q_{At} \tag{6}$$

where $q_{At}$ is the quantity of the risky asset purchased.

The asset's equilibrium price $P_t^*$ is given by:

$$P_t^* = \frac{\alpha_{At} W_{At} + \alpha_{Bt} W_{Bt}}{N}, \qquad s.t.\ \alpha_{it} \in (0,1], \forall i \tag{7}$$

where $W_{At}$ and $W_{Bt}$ are the total capitals of Trader A and Trader B, respectively, net of information costs. As mentioned above, the allocation of the risky asset that Trader A receives is proportional to Trader A's demand in the total market demand:

$$q_{At} = N * \frac{\alpha_{At} W_{At}}{\alpha_{At} W_{At} + \alpha_{Bt} W_{Bt}} \tag{8}$$

Combining (6)-(8), Trader A's profit is:

$$\pi_{At} = (F_t - \frac{\alpha_{At} W_{At} + \alpha_{Bt} W_{Bt}}{N}) * N * \frac{\alpha_{At} W_{At}}{\alpha_{At} W_{At} + \alpha_{Bt} W_{Bt}} =$$

$$= \left(\frac{F_t * N * \alpha_{At} W_{At}}{\alpha_{At} W_{At} + \alpha_{Bt} W_{Bt}}\right) - \alpha_{At} W_{At} \tag{9}$$

It is easy to verify that the second-order conditions on (9) do not allow Trader A to find a globally optimal $\alpha_{At}$. However, for $\alpha_{Bt} \to 0$, $\pi_{At}$ is reduced to:

$$\lim_{\alpha_{Bt} \to 0} \pi_{At} = F_t * N - \alpha_{At} W_{At} \tag{10}$$

And hence, the profit-maximizing strategy for Trader A, when $\alpha_{Bt} \to 0$, is to set $\alpha_{At} \to 0$ as well. That is, Traders A and B can achieve a mutually beneficial outcome by coordinating on a low value of $\alpha_{it}$. But for any value of $\alpha_{Bt}$ *not* arbitrarily close to zero, the profit-maximizing value of $\alpha_{At}$ depends on the specific values of $F_t$, $N$ and $\alpha_{Bt}$. Intuitively, whether Trader A can increase her profit by *decreasing* $\alpha_{At}$ depends on how much this reduces the equilibrium price $P_t^*$ relative to reducing the asset allocation $q_{At}$ that Trader A receives, given the specific values of $F_t$, $N$ and $\alpha_{Bt}$.

Since the traders in our model know neither their competitors' strategies nor even how many other traders they are in competition with, they can only try to learn an optimal strategy with respect to $\alpha_{it}$ by trial and error. The learning algorithms employed by the traders are described in detail in the next section.

*2.4 Learning algorithm*

The traders in our model are adaptive and seek to learn an optimal strategy both with respect to the information acquisition and the aggressiveness of trading via their own interaction with the market. Learning from feedback is typically modelled both in economics and computer science by reinforcement learning algorithms. Although various reinforcement learning algorithms exist, we opt for the so-called multiarmed bandit (MAB) family of algorithms for the following reasons. First, as [Han 2022] and [Cartea et al., 2022] point out, in financial markets traders typically do not have a good model of the environment they act in (e.g. the number of other agents in the market, their strategies, risk preferences, endowments etc.). Moreover, the traders do not even have reliable information regarding the market demand, due to either hidden orders being used in public trading venues [Boulatov & George, 2013], which according to [Bartlett & O'Hara, 2024] may currently represent up to 45% of equity trading volume, or due to trading in private venues, the so-called "dark pools" [Buti et al. 2017]. Therefore, there is no straightforward way to define the "states" which traders can find themselves in. Such states, however, need to be defined to use more advanced reinforcement learning algorithms. Second, by choosing relatively simple MAB learning algorithms we keep the complexity of our model to a minimum. More complex learning algorithms, such as e.g.

Q-learning with states or various implementations of deep reinforcement learning, would require us to set multiple parameters in an ad hoc fashion. In addition to the lack of empirical basis for these parameters, their multitude also makes the attribution of results more difficult. In contrast, the MAB learning algorithm we employ, and particularly its Upper Confidence Bound (UCB) implementation, has only 1 parameter.

The UCB algorithm used by the traders in our model has two elements: the element estimating the "value" of an action (the return of a given strategy in our case) and the element balancing the exploitation of the actions that have been found advantageous with the exploration of actions that have not been sufficiently experimented with.

In our model the trader $i$'s estimate of the return provided by the strategy $k$ at period $t$ is given by:

$$Q_{ik} = \frac{R_1 + R_2 + \cdots + R_{n-1}}{n-1} \tag{11}$$

where $n$ is the number of times the strategy $k$ has been tried by the trader up to the current period and $R_n$ are the trader's own empirically observed returns while using the strategy $k$.

Having updated the estimates $Q_{ik}$ after each round of trading, the trader $i$ chooses the strategy $k$ for the next period according to the following criterion:

$$k_{it} = \arg\max_k \left[ Q_{ik} + c \sqrt{\frac{\ln t}{N_{kt}}} \right], \qquad k \in K \tag{12}$$

where $N_{kt}$ is the number of times the strategy $k$ has been tried up to the time point $t$ and $c$ is a constant determining the weight of the "exploration term" versus the "exploitation term".

Although [Cartea et al., 2022] point out that the UCB algorithm is not the best of MAB algorithms at dealing with adversarial environments, its advantage compared to other implementations of MABs is that it balances exploitation and exploration in a principled way, whereby non-optimal actions are explored the more frequently, the less they have been tried in the past. Other MAB implementations, such as e.g. the so-called ε-greedy algorithms, do not have this property (see [Cartea et al., 2022] and [Sutton & Barto, 2018] for a more extensive discussion).

The timeline of a typical simulation round is presented in Figure 1. The parameters of the simulations are shown in Table 1. We discretize the action space with respect to α into the set

{1, 0.75, 0.5, 0.25, 0.01}, as learning the values of a continuous action space would require the use of an approximation algorithm, such as e.g. deep neural nets. Since, as mentioned above, this would necessitate an introduction of a large number of additional parameters for which we do not have a clear economic justification, we stick to a discrete action space.

Similar to [Pastushkov 2024], we run computational experiments for various levels of the information cost $C$ to be able to compare our results with the competitive case as well as to see how different costs $C$ affect the outcomes of our model. To enhance the comparability with [Pastushkov 2024], the number of simulation steps is set to T=2500 (see Table 1). For each computational experiment we run 50 stochastic simulations.

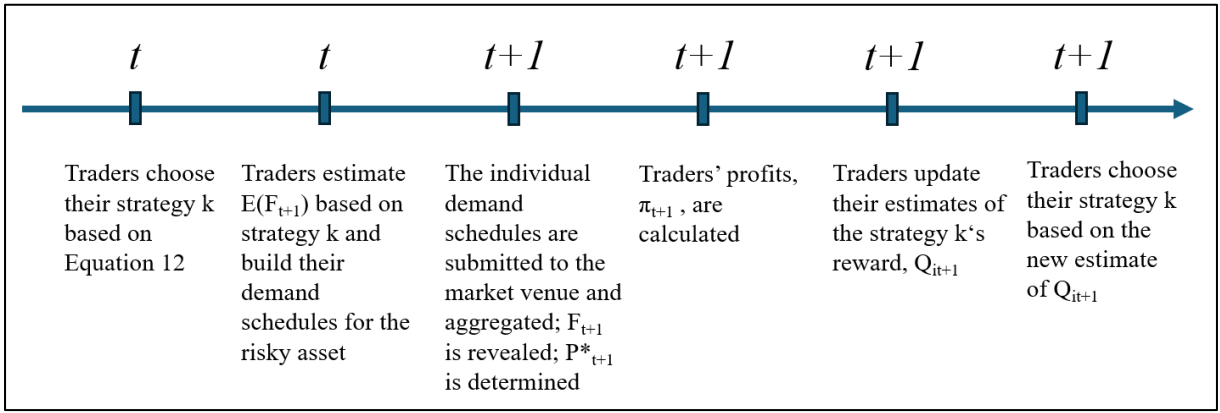

**Fig 1.** Timeline of a simulation round

| Parameters | Values |
|---|---|
| Cost of information, C | {0, 0.02, 0.04, 0.06, 0.08, 2, 4, 6, 8, 10} |
| Number of traders, M | 100 |
| Number of shares, N | 1000 |
| Initial capital of trader i | U(0,1000] |
| Initial value of the payoff, F0 | 30 MU |
| Risk-free rate, $r_f$ | 0.01/252 |
| Average growth rate of the payoff, $\mu$ | 0.1/252 |
| St. deviation of the payoff value growth, $\sigma$ | 0.01 |
| Number of simulation steps, T | 2500 |
| Exploration constant, c | 0.001 |
| Share of capital invested in the risky asset, $\alpha$ | {0.01, 0.25, 0.5, 0.75, 1} |

**Table 1**. Simulation parameters

## Section 3. Results and discussion

*3.1 Market efficiency*

Since in our model the risky asset provides a single payoff, we define an informationally efficient market as the one in which the risky asset's market price corresponds to the value of the payoff discounted at the risk-free rate. We refer to any deviation of the equilibrium price $P_t^*$ from the discounted payoff $F_t/(1+r_f)$ as mispricing. Since mispricing can refer to both over- and underpricing, we calculate the absolute mispricing $\varepsilon_t$ for every simulation period, so that periods of over- and underpricing do not cancel each other out:

$$\varepsilon_t = \left| \frac{P_t^*}{F_t/(1+r_f)} - 1 \right| \qquad (13)$$

We then calculate the average mispricing over the whole duration of a simulation run:

$$\bar{\varepsilon} = \frac{\sum_{t=0}^{T} \varepsilon_t}{T} \qquad (14)$$

Finally, we average $\bar{\varepsilon}$ across the 50 simulation runs for every value of the information cost $C$. The results are presented in Figure 2. For comparison, we reproduce the results of [Pastushkov 2024] for the case where the traders are purely competitive and can only choose whether to become informed or uninformed, that is, the set of strategies available to them is $K = \{Inf_{\alpha=1}, Uninf_{\alpha=1}\}$.

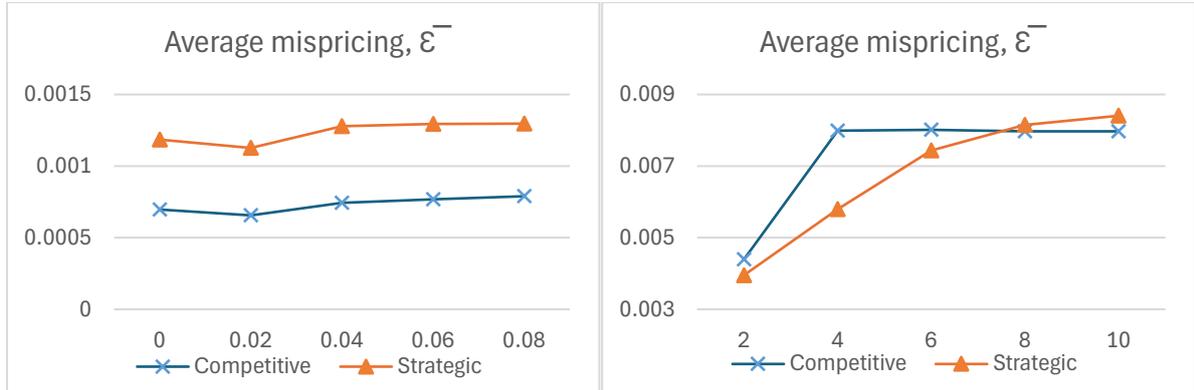

A.            B.

**Fig. 2** Average mispricing for the strategic and the competitive case. Panels A and B report the results for the low and the high levels of the information cost *C*, respectively. The information costs are shown on the *x* axis.

Comparing Panels A and B of Figure 2 one can observe that the mispricing increases for higher information costs *C*, and this happens both for the competitive and the strategic case. This result is hardly surprising, given both the predictions of [Grossman & Stiglitz, 1980] and [Kyle 1989] as well as the previously reported simulation results of [Pastushkov 2024] for the competitive case. However, a surprising pattern emerges when one compares the results of the strategic to the competitive case. While for the low information cost regime (Panel A), the results conform to the prediction of [Kyle 1989] in that the traders choosing their demand levels strategically lead to a higher mispricing than their purely competitive counterparts, for the high information cost regime (Panel B) the pattern is pretty much reversed, as the strategically acting traders lead the market to a *lower* average mispricing, and hence *higher* informational efficiency. This is in sharp contrast to the theoretical results of [Kyle 1989] who predicted that a market with strategic traders must be less efficient than one with competitive traders, ceteris paribus, regardless of the information costs. Before examining the causes of these results in Sections 3.3 and 3.4, we report some further aggregate properties of the simulated market.

*3.2 Mispricing and wealth dynamics*

Following [Pastushkov 2024], in Figures 3-8 we show the moving average mispricings with a lookback window of 300 periods, averaged across the simulation runs, for several selected levels of the information cost $C$. In each of the Figures 3-8, Panel A presents the "competitive" case whereas Panel B presents the "strategic" one. The Figures 3-8 also report the aggregate capitals of the informed and the uninformed traders, whereby moving averages with the same lookback window as for the mispricing are taken, and individual histories are averaged across the simulation runs for a given information cost $C$.

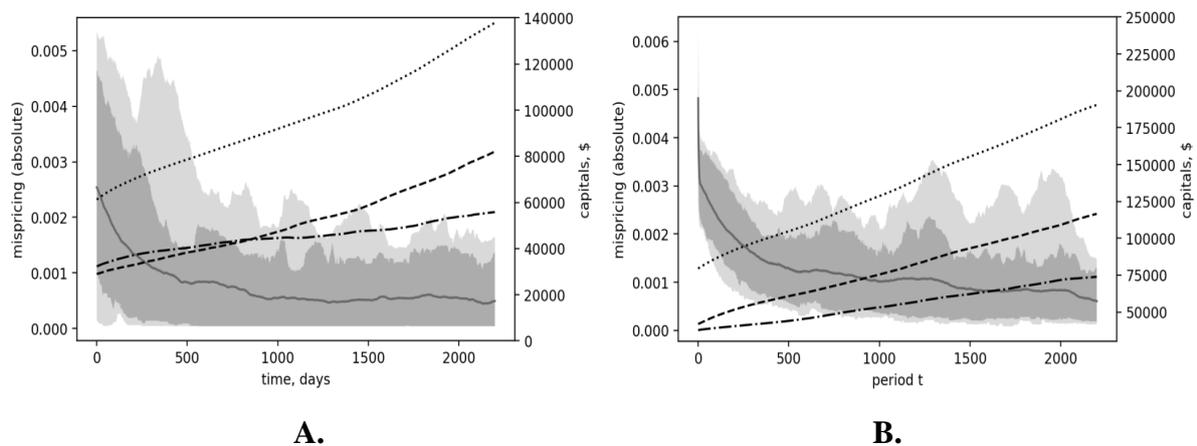

A.    B.

**Fig. 3** $C=0$, the solid grey line shows the absolute mispricing, the dashed black line shows the aggregate capital of the informed traders, the dash-dot black line shows the capital of the uninformed, the dotted line shows the sum of the informed and the uninformed traders' capitals.

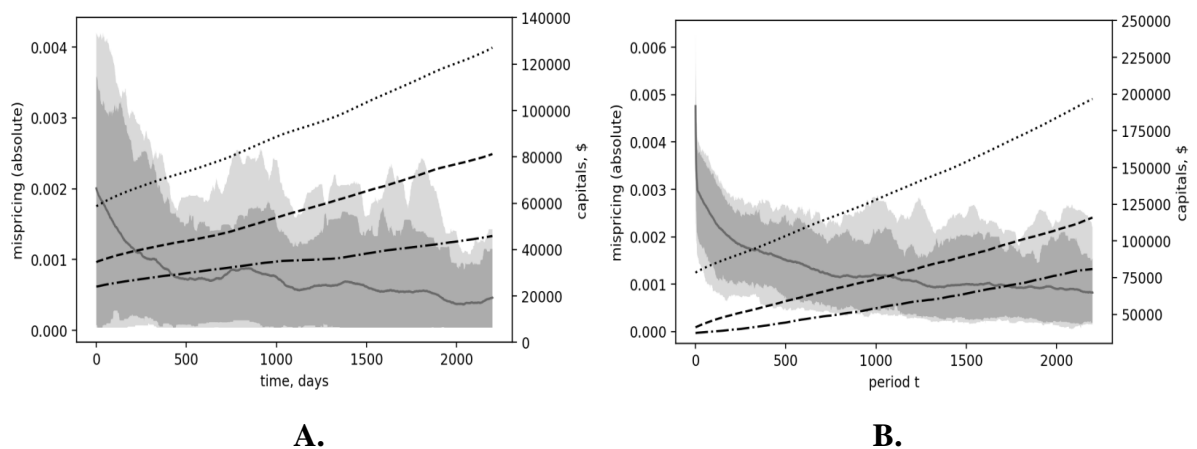

A.    B.

**Fig. 4** $C$=0.04, the solid grey line shows the absolute mispricing, the dashed black line shows the aggregate capital of the informed traders, the dash-dot black line shows the capital of the uninformed, the dotted line shows the sum of the informed and the uninformed traders' capitals.

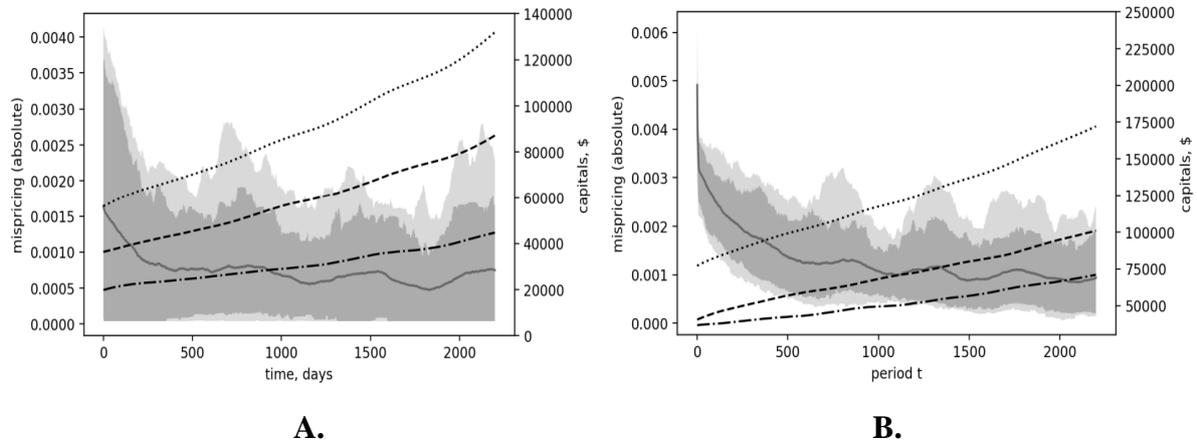

A.  B.

**Fig. 5** $C$=0.08, the solid grey line shows the absolute mispricing, the dashed black line shows the aggregate capital of the informed traders, the dash-dot black line shows the capital of the uninformed, the dotted line shows the sum of the informed and the uninformed traders' capitals.

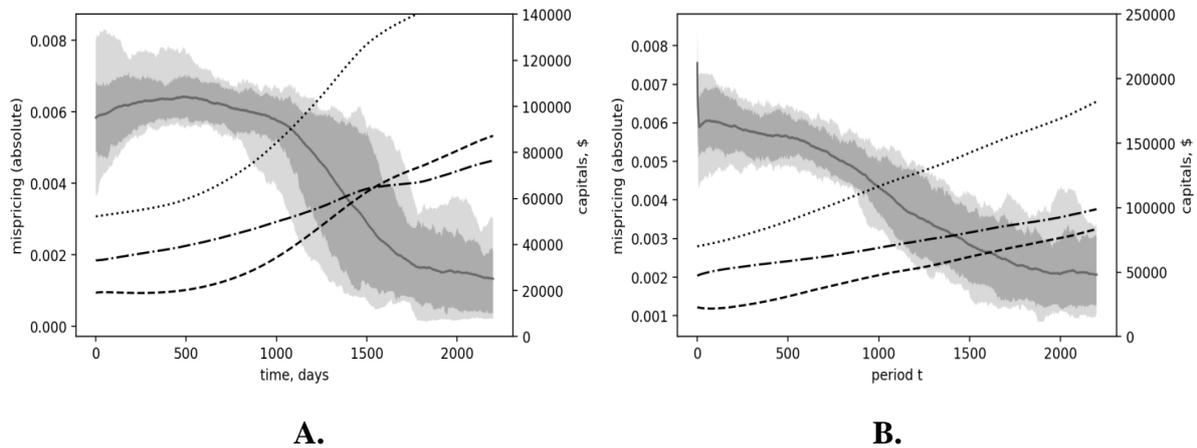

A.  B.

**Fig. 6** $C$=2, the solid grey line shows the absolute mispricing, the dashed black line shows the aggregate capital of the informed traders, the dash-dot black line shows the capital of the uninformed, the dotted line shows the sum of the informed and the uninformed traders' capitals.

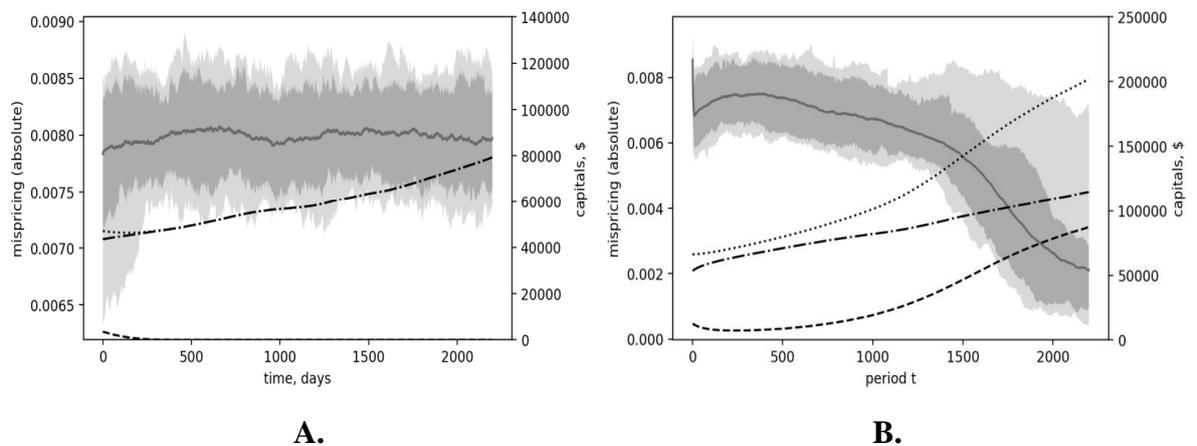

A.  B.

**Fig. 7** $C$=4, the solid grey line shows the absolute mispricing, the dashed black line shows the aggregate capital of the informed traders, the dash-dot black line shows the capital of the uninformed, the dotted line shows the sum of the informed and the uninformed traders' capitals.

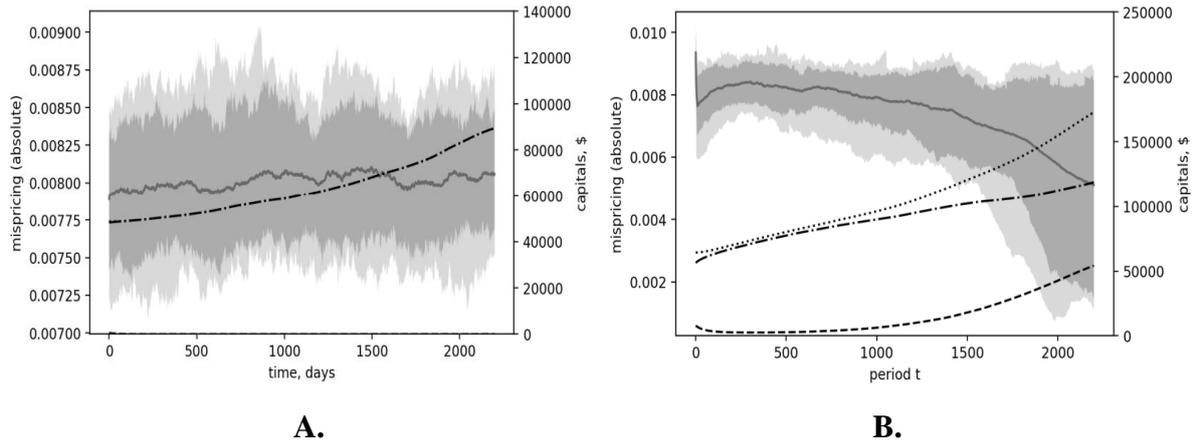

         **A.**               **B.**

**Fig. 8** $C$=6, the solid grey line shows the absolute mispricing, the dashed black line shows the aggregate capital of the informed traders, the dash-dot black line shows the capital of the uninformed, the dotted line shows the sum of the informed and the uninformed traders' capitals.

Examining Figures 3-8 we observe several notable patterns. First, for the case of zero information costs (Figures 3A and 3B), the initial dominant wealth share of the uninformed traders that was observed for the competitive case and reported in [Pastushkov 2024] is not observed for the strategic case. The reason for this is simple: since in the competitive market the traders can only choose between two strategies (informed and uninformed) and their reward estimates $Q_{ik}$ are calculated based on the percentage return on capital, when the cost of information is zero the estimates $Q_{ik}$ that the informed traders make are equalized, which leads to a synchronization of their learning. This synchronization, in turn, leads to a higher probability that *all* traders in the market would simultaneously choose the uninformed strategy in the process of exploration, causing the average mispricing to increase. When traders learn asynchronously, such situations of uniform selection of the uninformed strategy are extremely unlikely to occur. In the strategic case, however, even if the information cost is zero (Figure 3B) the traders' learning does not synchronize since they have more strategies to explore, and hence the aforementioned negative effect on the market efficiency disappears.

Second, one can observe that for the high levels of $C$ the informed traders control a sizeable share of the market wealth in the latter periods when strategic trading is allowed (Figures 6-8B), unlike in the competitive case, in which the informed traders are driven out of the market (Figures 7-8A). This appears to be the proximate cause of the improvement of the average market efficiency in the high informational cost regime (Figure 2B) under strategic trading. The greater share of wealth controlled by the informed traders leads to them driving the market prices closer to the "fundamental" payoff values, thus increasing efficiency. However, strategic trading does not improve the average market efficiency in the low information cost regime. Intuitively, when strategic trading is introduced, market efficiency is affected by two opposite effects: on the one hand, the informed traders, by utilizing their market power, are able to accumulate a larger wealth share and hence drive the market to higher efficiency, on the other, if traders exploit their effect on the equilibrium price, they strategically constrain their demand such that the asset's price is driven further away from the payoff value compared to the purely competitive case. In the next section we take a closer look at the degree to which the market in the aggregate is affected by this strategic constraining of liquidity for various information cost regimes. We refer to this phenomenon as simply "non-competitive" pricing, as opposed to the term "supra-competitive" used, among others, by [Cartea et al., 2022] and [Calvano et al., 2020]. This is because in our model the asset's price can only deviate from the "fundamental" payoff *downwards* as a result of strategic constraining of liquidity, whereas in [Calvano et al., 2020] and in [Cartea et al., 2022] non-competitive market makers charge a spread *above* the "fair" competitive price, and hence the price becomes "supra-competitive".

*3.3 Non-competitive pricing*

We follow [Calvano et al., 2020] in reporting relative values of non-competitive pricing. Hereby our lower and higher bounds, which serve us as benchmarks, are given by the cases where *all* traders choose the maximally liquidity-constraining (α=0.01) and the maximally "competitive" (α=1) strategy, respectively. That is, our Δ-statistic is given by:

$$\Delta_t = \frac{P_t^* - P_{t,\alpha_i=0.01}}{P_{t,\alpha_i=1} - P_{t,\alpha_i=0.01}}, \qquad \forall i \in M \tag{15}$$

where $P_t^*$ is the actual equilibrium price observed in the simulations, and $P_{t,\alpha_i=0.01}$ and $P_{t,\alpha_i=1}$ are *potential* equilibrium prices that *would have been observed* if all traders had chosen α=1 and α=0.01, respectively. Similar to Figures 3-8, we calculate moving averages of $\Delta_t$ with a lookback window of 300 periods to remove the excessive jaggedness of the graphs and average the resulting graphs across the simulation runs. The results are presented in Figures 9-14.

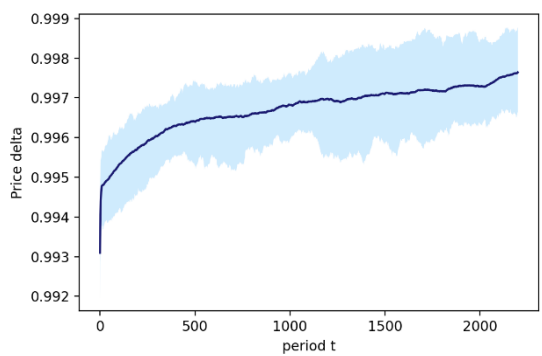
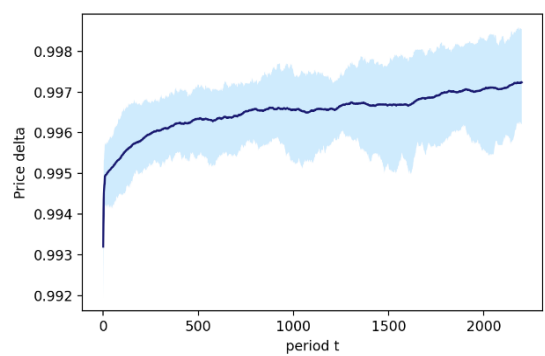

**Fig. 9**   Average $\Delta_t$, *C*=0                    **Fig. 10** Average $\Delta_t$, *C*=0.04

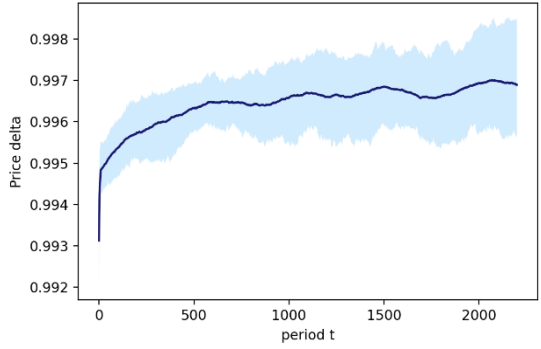

**Fig. 11** Average $Δ_t$, $C$=0.08

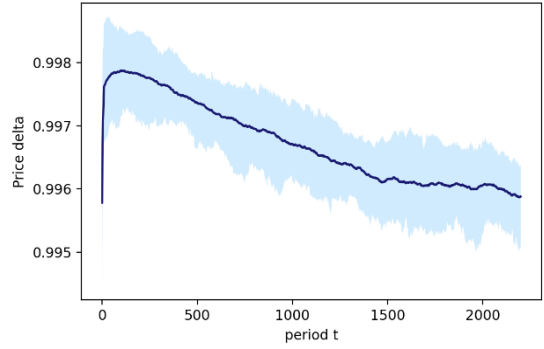

**Fig. 12** Average $Δ_t$, $C$=2

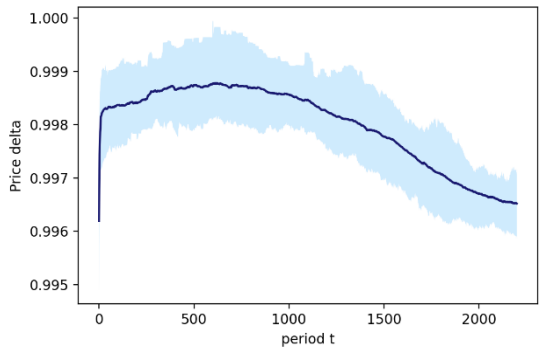

**Fig. 13** Average $Δ_t$, $C$=4

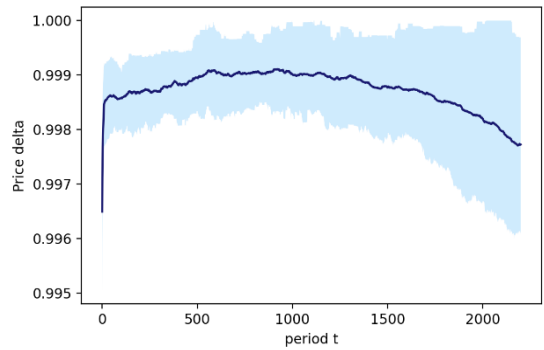

**Fig. 14** Average $Δ_t$, $C$=6

One can observe across the different levels of *C* that although the deviations from the purely competitive pricing (the upper bound of 1.0 in Figures 9-14) are small, $Δ_t$ *never* reaches the purely competitive level. This is true both for the averaged across the simulation runs $Δ_t$ (represented by the blue line), as well as for its $10^{th}$ and $90^{th}$ percentiles shown by the blue margins surrounding the blue line. Notably, the pricing is more competitive in the high informational cost regime (Figures 12-14), as the blue line on average is closer to the upper bound.

We conclude that when traders have an opportunity to strategically constrain their liquidity provision in order to improve their returns, they (in the aggregate) learn to make use of this opportunity. The aggregate effect, however, is rather small and it is smaller for the high information cost regime.

At this point it is worth highlighting that our results contradict [Cartea et al., 2022] who show that a modest number (5 to 7) of competing agents using a UCB algorithm is enough to eliminate non-competitive pricing completely in their model. In our model, non-competitive pricing persists despite there being a relatively large number of competing, UCB-driven agents (M=100). There are, however, important differences between the model of [Cartea et al., 2022] and the one presented here. First, ours is a model of an order-driven market in which every trader can act as a liquidity provider, whereas in [Cartea et al., 2022] the market is quote-driven and the prices are set by specialized market makers.[6] Second, in our market the supply of the risky asset N is provided by a liquidity-seeking representative trader, who sends a market order which is insensitive to the price. In [Cartea et al., 2022] on the other hand, the liquidity-seeking trader who requests quotes from the market makers has a reservation price, and therefore, their demand is sensitive to the quotes offered. The trader may not trade at all if none of the market makers offers an acceptable quote. Such situations are impossible in our model, in line with the actual behavior of market orders. Taken together, these substantial differences may account for the different results with respect to the persistence of non-competitive pricing reported by [Cartea et al., 2022] and observed in our model. In any case, [Cartea et al., 2022] point out that their results are only valid for the model of market interactions they assume, and different market structures may give rise to different results, which is indeed what happens in our model.

With regard to the average market efficiency, it appears that although our market indeed shows a degree of non-competitive pricing when traders are strategic across all examined levels of *C*, this effect leads to worse average efficiency only in the low information cost regimes. In the high information cost regimes, the net effect of the non-competitive pricing and the higher share of capital controlled by the informed traders is positive, leading to improved market efficiency, compared to the purely competitive case.

*3.4 What strategies do traders learn?*

While the net-positive effect of non-competitive pricing and a higher level of informed trading has been found to be the immediate cause for the higher average efficiency of the "strategic" market in the high information cost regime, it is interesting to examine *why* the informed traders are more likely to survive in the high information cost regime when strategic trading is introduced. We therefore examine what strategies the traders learn. More specifically, our goal is to examine the relationship (if any) between information acquisition and strategic trading. Do both the informed and the uninformed traders take advantage of the opportunity to strategically constrain their liquidity provision in order to suppress the equilibrium price? Or is one type of traders more likely to trade strategically than the other? How does the introduction of strategic trading enhance the performance of the informed traders in the high information cost regimes?

---

[6] It can be argued that while the model of [Cartea et al., 2022] is a good model of actual OTC markets, such as the FX and the interest rate market, ours more closely models centralized markets for equities and equity derivatives.

The analysis is complicated somewhat by the fact that information acquisition (as is strategic trading) is endogenous in our model, and therefore there is no straightforward way to discretely divide the traders into the informed and the uninformed group. Instead, we build scatterplots, showing on the x-axis the number of periods in a given simulation run that a given trader chose to be informed (uninformed), and on the y-axis the percentage of times that the trader chose the competitive liquidity provision strategy (α=1) while being informed (uninformed). The results are shown in Figures 15-18.

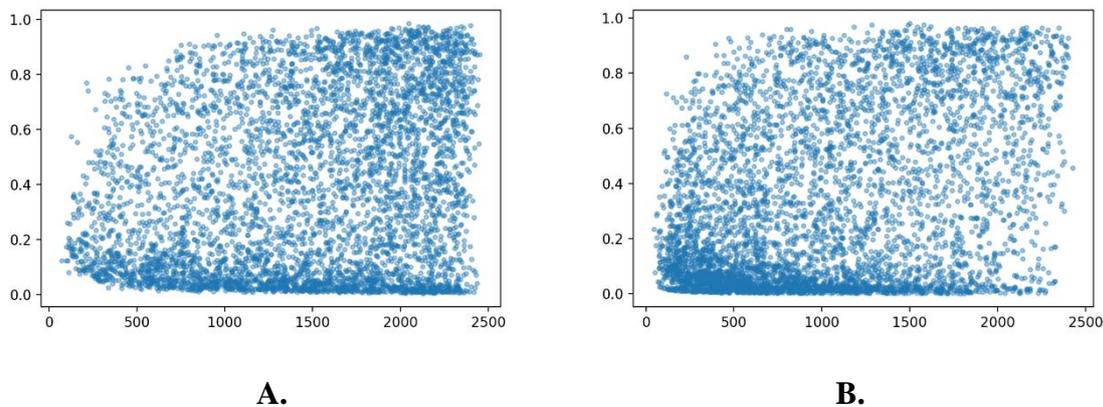

A.                  B.

**Fig. 15** *C*=0. *X* axis of Panel A (Panel B) shows the number of rounds a given trader chose to be informed (uninformed). *Y* axis shows the % of rounds a given trader chose the fully competitive strategy (α=1) while being informed (uninformed).

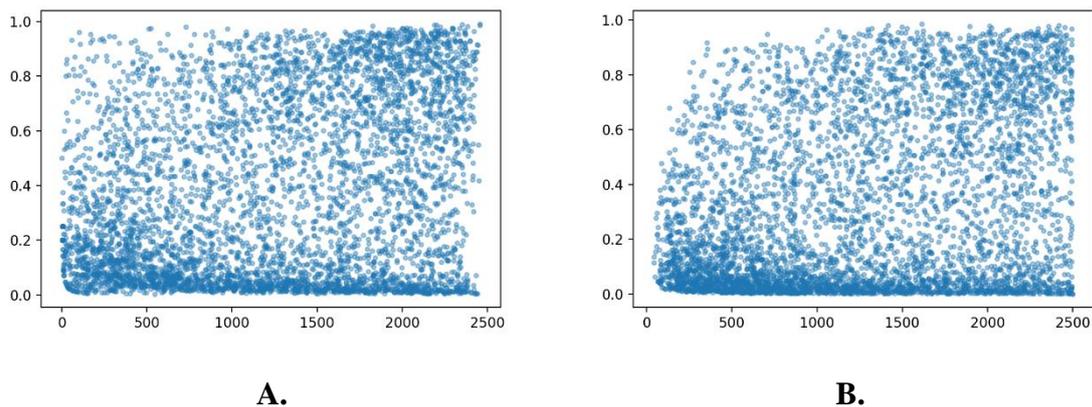

A.                  B.

**Fig. 16** *C*=0.08. *X* axis of Panel A (Panel B) shows the number of rounds a given trader chose to be informed (uninformed). *Y* axis shows the % of rounds a given trader chose the fully competitive strategy (α=1) while being informed (uninformed).

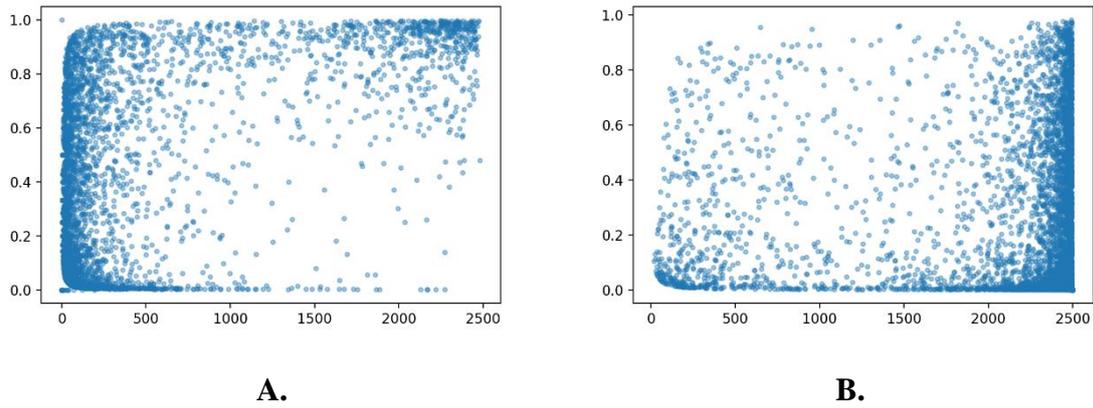

**A.**                                          **B.**

**Fig. 17** *C*=2. *X* axis of Panel A (Panel B) shows the number of rounds a given trader chose to be informed (uninformed). *Y* axis shows the % of rounds a given trader chose the fully competitive strategy ($\alpha$=1) while being informed (uninformed).

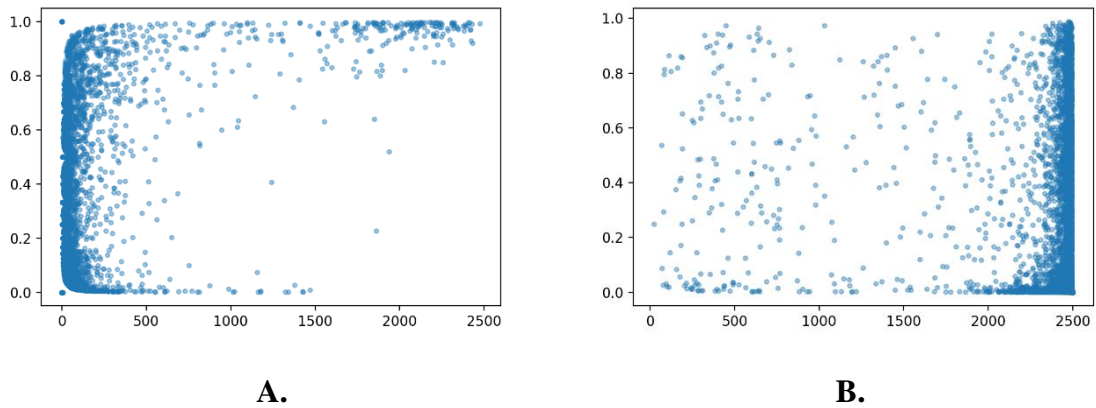

**A.**                                          **B.**

**Fig. 18** *C*=4. *X* axis of Panel A (Panel B) shows the number of rounds a given trader chose to be informed (uninformed). *Y* axis shows the % of rounds a given trader chose the fully competitive strategy ($\alpha$=1) while being informed (uninformed).

We do not show scatterplots for all levels of *C*, as the ones shown in Figures 15-16 and 17-18 are highly representative of the low and high information cost regimes, respectively. As Figures 15A-18A show on the x-axis the number of times a given trader was *informed*, one can think of the dots on the right-hand side of the scatterplot as representing the informed traders, and the dots on the left-hand side – the uninformed ones. The opposite is true of the Figures 15B-18B: since the x-axis shows the number of times a given trader chose to be *uninformed*, the dots to the right of the middle represent the uninformed traders, and to the left – the informed ones.

One can observe that for the cases of low information costs (Figures 15-16) the dispersion of how often traders choose to trade competitively is rather high. This is true both of the informed and the uninformed traders. As could be expected, the majority of traders tend to be informed in the low information cost regime, as is evident by a higher concentration of dots on the right-hand side of Figures 15A-16A and on the left-hand side of Figures 15B-16B. Noticeable large clusters of dots in the bottom left corners of Figures 15B-16B signal that when traders who generally tend to be informed experiment with the uninformed strategy, they also tend to constrain their liquidity provision (percentage of time they choose the "competitive" liquidity provision, shown on the y-axis, is between 0 and 0.2). When they exploit the informed strategy, however, they are not as likely to use liquidity-constraining strategies, as the top right corners of Figures 15A-16A also show significant concentrations of dots. The uninformed traders, on the other hand, tend to constrain liquidity provision when they experiment with the informed strategy (see noticeable clusters of dots in the bottom left corners of Figures 15A-16A). There is also a notable difference in the behavior of the uninformed traders for the *C=0* and *C*=0.08: while in the former case they tend to trade competitively when using the uninformed strategy (top right corner of Figure 15B), in the latter case, with increasing information costs, they are quite likely to constrain liquidity while exploiting the uninformed strategy (see the concentration of dots in the bottom right corner of Figure 16B).

When we consider the high informational cost regime, the tendency of the informed traders to trade competitively when informed and constrain liquidity when uninformed becomes much more apparent, as one can observe pronounced clusters of dots in the top right corners of Figures 17A-18A and bottom left corners of Figures 17B-18B. The liquidity provision by the uninformed traders is much more dispersed. However, as somewhat larger concentrations of dots seem to appear for them in the bottom parts of the scatterplots, one can conclude that they tend to constrain liquidity both when using the informed and the uninformed strategy.

Returning to the question asked in the beginning of this section, it appears that the informed traders are better able to accumulate capital in the high informational cost regimes due to learning to strategically constrain their liquidity provision when using the uninformed strategy. At the same time, the uninformed traders are only slightly more likely to constrain liquidity than to trade competitively in the high informational cost regimes, which means that the introduction of strategic trading leads to only a slight degree of non-competitive pricing in the high informational cost regimes. Taken together, the slightly non-competitive pricing of the asset and the larger share of trading driven by the informed traders leads to a net improvement of market efficiency for the high informational cost regimes, compared to the purely competitive case (see Figures 2A and 2B).

*3.5 Pseudo-collusion*

Finally, we want to examine whether the traders in our model behave in a pseudo-collusive way, as defined by [Cartea et al., 2022]. Obviously, since the traders do not observe each other's actions with regard to both information acquisition and liquidity provision, *tacitly* collusive effects cannot arise, as tacit collusion requires a possibility of punishment for deviations from the collusive behavior. But pseudo-collusion requires only that agents learn to perform actions that are not individually optimal for them, in order to achieve a cooperative outcome. Since our model does not allow us to analytically derive a globally individually optimal strategy (see Section 2.3), we examine this question empirically by checking in every period and for every agent whether there are actions that would have led to a higher profit for her if she deviated from her *learned* action. That is, we recalculate the equilibrium market price and the resulting profit of every agent in every period based on alternative strategies that the agent *could have chosen*, and check whether the strategies *actually chosen* by the agent were indeed the individually optimal ones. If that is not the case, one can speak of pseudo-collusion, or cooperative behavior, as then the agents choose the actions that are not individually optimal, but collectively lead to beneficial outcomes.

Before reporting the statistics, however, we want to make sure that the learning algorithms that the agents use have converged. It has been found in the course of the simulation experiments that on average after ca. 8000 periods ca. 80% of the agents stick to a single strategy (see Figure 19 for some representative graphs of aggregate convergence). We therefore run the simulations for 2000 more periods after the initial 8000 steps to obtain the results on the agents' behavior after the convergence is largely achieved. Subsequently, taking the final 2000 simulation periods, we find the agents who stick to a single given strategy 80% of the time or more. These constitute our group for which we check whether the stable strategy they use is the individually optimal one. If they use the individually optimal strategy at least 50% of the time (i.e. 1000 of the last 2000 periods or more), we count an agent as having converged to the optimal strategy.

Finally, we average the numbers of the converged agents and the percentages of them who converged to the optimal strategy across the 50 simulation runs for every level of the information cost *C*. The percentage of the agents who converged to the optimal strategy is hereby weighted by the total percentage of the agents who converged in an individual simulation run. The results are reported in Table 2.

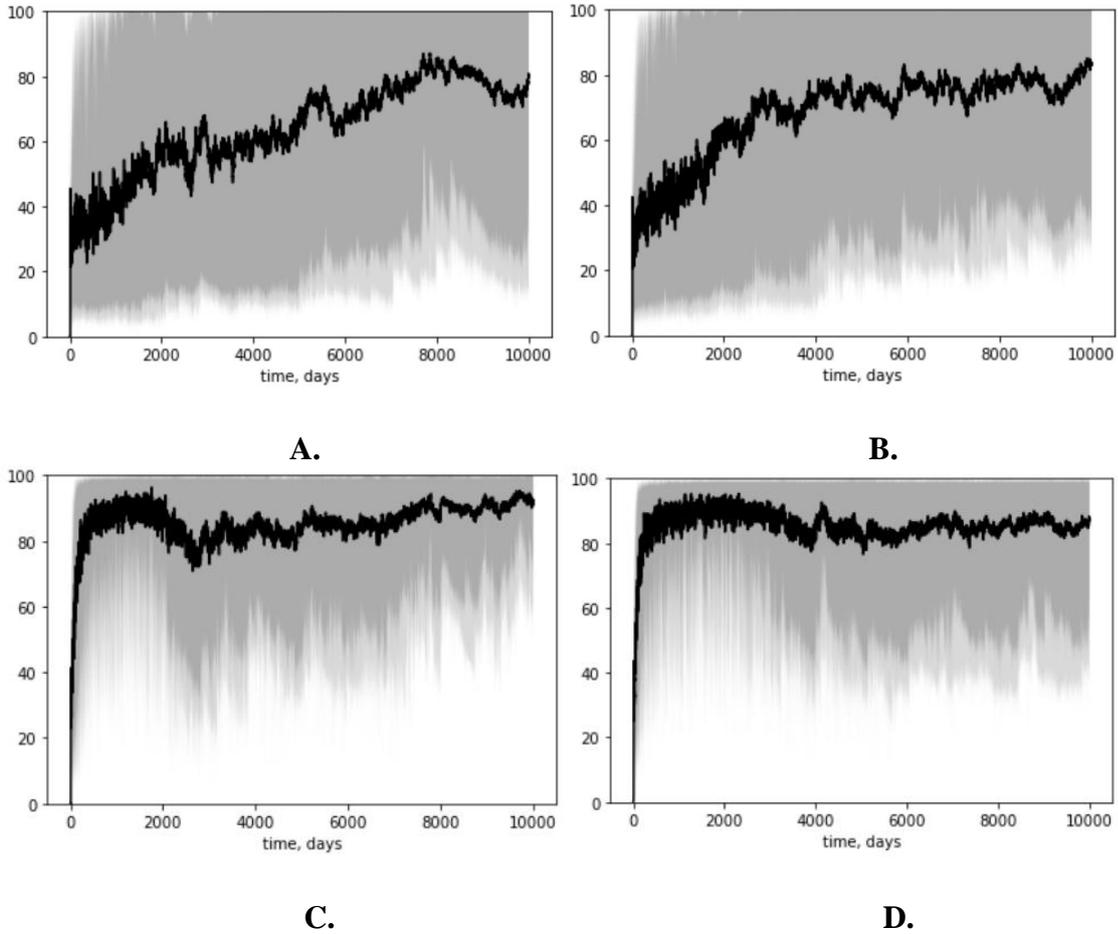

**Fig. 19** Number of agents sticking to a single strategy k, averaged across simulation runs. The grey margins surrounding the black line represent the 10[th] and 90[th] percentiles. Panels A, B, C and D show the results for the information costs of C={0, 0.08, 6, 8}, respectively.

| Cost of information, C | 0 | 0.02 | 0.06 | 0.08 | 2 | 4 | 6 | 8 | 10 |
|---|---|---|---|---|---|---|---|---|---|
| Average number of converged agents | 56.2 | 49.9 | 48.9 | 54.8 | 80.2 | 82.1 | 76.0 | 68.0 | 73.0 |
| St. deviation of converged agents | 29.5 | 29.3 | 31.2 | 27.6 | 18.8 | 19.5 | 21.6 | 23.9 | 21.8 |
| Average percentage of converged agents using the optimal strategy | 76% | 30% | 36% | 36% | 57% | 61% | 63% | 57% | 58% |

**Table 2**

One can observe that the share of agents whose learning has converged is noticeably higher for the high information cost regimes (*C* of 2 to 10), and the dispersion of this number is also smaller. However, the highest percentage of agents who have *converged to the optimal strategy* is observed for the information cost of *C*=0. For the high information cost regime, these percentages are also substantial, reaching a maximum of 63% for *C*=6.

The lowest average percentages of agents who converged to the optimal strategy are observed for the low, but non-zero information costs. In these regimes, just above a third of the agents whose learning algorithms have converged tend to use the optimal strategy. This means that for these levels of *C* the probability of agents exhibiting pseudo-collusive behavior is the highest, i.e. the majority of the agents has converged to using a suboptimal strategy to improve the collective outcome. This finding matches our earlier results showing that the degree of non-competitive pricing is more pronounced for the low levels of *C*, and sheds light on why the introduction of strategic trading tends to worsen the average market efficiency for the low levels of *C*.

**Section 4. Conclusion and further research directions**

In the present study we examined how an introduction of strategic trading (i.e. an opportunity for traders to strategically constrain liquidity provision to suppress the market price of a risky asset) affects the market efficiency, compared to the purely competitive case. Hereby we used a recent study by [Pastushkov 2024] as a benchmark providing the results for a competitive market. We found that the predictions of the static model of [Kyle 1989] (i.e. that a market with strategic traders is necessarily less efficient than a purely competitive one) hold for some levels of the information cost, and do not hold for others. We examined the causes of these outcomes and found that for some information cost regimes, the net effect of non-competitive pricing and a higher probability of survival for the informed traders is ultimately positive for market efficiency. I.e. in these regimes the informed traders driving the market price closer to the "fundamental" value have a more pronounced effect than the inefficiencies caused by strategically constrained liquidity provision.

Furthermore, this study provides further evidence on the ability of a large number of independent traders learning optimal strategies by MAB algorithms to coordinate on non-competitive pricing via constrained liquidity provision. Contrary to the results of [Cartea et al. 2022] for an OTC market with specialized market makers, we find that in a centralized market in which every trader can act as a strategic liquidity provider, even a relatively large number of traders [M=100] can coordinate on non-competitive pricing, thus improving their own profits and reducing market efficiency. It is worth highlighting, however, that the results presented here are valid only for the particular market model assumed, and the outcomes may vary for other market structures.

In our model we assumed that the traders use *stateless* MAB learning algorithms, as in modern financial markets a high share of demand and supply is latent. It would be interesting to examine how the results of our model might be affected by an alternative assumption of (partial) observability of the market demand by traders. This question is left for future work.